\documentclass[letter,english]{article}
\usepackage{emulateapj,onecolfloat,graphics}

\begin{document}

\twocolumn
[
\title{The distribution of mass around galaxies}
\author{Martin White, Lars Hernquist, Volker Springel}
\affil{Harvard-Smithsonian Center for Astrophysics, Cambridge, MA 02138}

\begin{abstract}
\noindent
We explore the distribution of mass about the expected sites of galaxy
formation in a high-resolution hydrodynamical simulation of a $\Lambda$CDM
cosmology which includes cooling, star-formation and feedback.
We show that the evolution of the galaxy bias is non-trivial and non-monotonic
and that the bias is stochastic.
We discuss the galaxy-mass cross-correlation function and highlight problems
with the interpretation of galaxy-galaxy lensing as due to extended dark
matter halos around ``typical'' galaxies.
Finally we show that the ``external shear'' around strong gravitational lenses
is likely to be closely aligned with the direction of the nearest massive
galaxy and to have a power-law distribution which can be predicted from the
galaxy auto-correlation function.
\end{abstract}
\keywords{cosmology: theory -- large-scale structure of Universe} ]

\section{Introduction}

The distribution of galaxies with respect to the total mass in the Universe
remains a central unsolved problem in cosmology, though one which is becoming
increasingly amenable to theoretical modeling and ever more constrained
observationally.  In particular, gravitational lensing allows us to probe the
spatial distribution of the mass in the Universe, for example around galaxies
and in clusters of galaxies.
In this paper we present predictions for the distribution of
mass around sites of galaxy formation using a hydrodynamic simulation of
structure formation which includes cooling, star-formation and feedback.
We believe such simulations can provide ab initio and robust predictions for
the statistical distribution of the sites of galaxy formation, even though
the detailed star formation rates and galaxy morphologies are sensitive to
uncertain ``sub-resolution'' physics.
As such, these simulations can be used to predict, from first principles,
the galaxy auto-correlation function, the galaxy-mass cross correlation
function (which is the key ingredient in galaxy-galaxy lensing studies) and
the mass auto-correlation function.

Originally regarded as an interesting, but not practically useful, prediction
of Einstein's theory of general relativity, gravitational lensing has become
a ``standard'' tool in observational cosmology.
Strong gravitational lensing has a long history, and weak lensing of
background galaxies by clusters of galaxies is now well established.
Just recently the weak lensing of galaxies by large-scale structure has
been observationally demonstrated
(van Waerbeke et al.~\cite{Ludo};
 Bacon et al.~\cite{BRE};
 Kaiser et al.~\cite{KWL};
 Wittman et al.~\cite{WTKDB};
 Maoli et al.~\cite{Maoli};
 Rhodes et al.~\cite{RRG}),
at about the level predicted by theories based on gravitational instability
in a cold dark matter (CDM) dominated universe.

The weak lensing observations probe the (projected) distribution of mass in
the Universe directly, without reference to the distribution of light.  This
provides us with direct constraints on the dark matter auto-correlation
function, for which theoretical predictions are very well developed both
analytically and numerically on Mpc scales.
By contrast, galaxy redshift surveys can give us detailed three-dimensional
information about the distribution of luminous matter, providing us with
a measure of the galaxy auto-correlation function and its evolution.
However, galaxy formation is still poorly understood, making interpretation of
the full information contained in the galaxy auto-correlation function
difficult.

Intermediate between these two are dark matter-galaxy cross-correlations.
These can be measured whenever the galaxy distribution is cross-correlated
with a tracer of the dark matter.
Some examples are galaxy-galaxy lensing (correlating galaxies and weak lensing
shear), foreground-background galaxy correlations and galaxy-QSO correlations
(correlating galaxies with weak lensing magnification).
All of these can be interpreted as a projection of the galaxy-mass correlation
function which we shall discuss.
To understand the galaxy-mass cross-correlation requires us to understand the
environments of galaxies, but not necessarily the detailed properties of the
galaxies themselves.

Another probe of the mass distribution around galaxies comes from strong
gravitational lensing.
Although the central regions of the primary lens galaxy, whose spatial
structure is not well modeled by current calculations, dominates the monopole
gravity in a strong lens, the quadrupole (and to a lesser extent the higher
poles) has a significant contribution from the external shear or tidal gravity
near the lens or along the line-of-sight.
All lensing models require this degree of freedom
(Barkana~\cite{Barkana96}; Keeton, Kochanek \& Seljak~\cite{Keeton97})
and in some cases the dominant source can be clearly identified with a nearby
galaxy.

The outline of this paper is as follows: we review the simulation we shall
use in \S\ref{sec:sim}, giving details of how we identify galaxies and their
parent halos.  The evolution of clustering in the simulation is discussed
in \S\ref{sec:clustering} and we introduce our primary tool, the correlation
function in \S\ref{sec:corrfn}.  The implications of our simulation for the
interpretation of galaxy-galaxy lensing is dealt with in \S\ref{sec:gglensing}.
Models of external shear for strong lensing are dealt with in
\S\ref{sec:strong}.  Finally in \S\ref{sec:conclusions} we summarize
our findings.

\section{The simulation} \label{sec:sim}

Throughout, we shall use a new simulation of the Ostriker \& Steinhardt
(\cite{OstSte}) concordance model, which has
$\Omega_{\rm m}=0.3$, $\Omega_\Lambda=0.7$,
$H_0=100\,h\,{\rm km}{\rm s}^{-1}{\rm Mpc}^{-1}$ with $h=0.67$,
$\Omega_{\rm B}=0.04$, $n=1$ and $\sigma_8=0.9$
(corresponding to $\delta_H=5.02\times 10^{-5}$).
This model yields a reasonable fit to the current suite of cosmological
constraints and as such provides a good framework for making realistic
predictions.

\begin{figure}
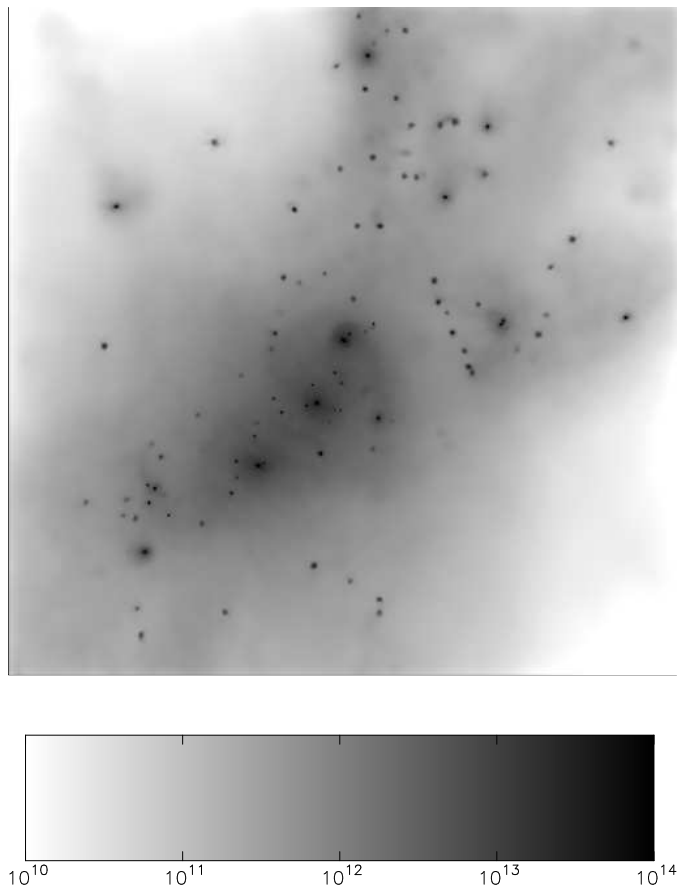

\begin{center}
\resizebox{3.5in}{!}{\includegraphics{gasmass.ps}}
\end{center}
\begin{center}
\resizebox{3.5in}{!}{\includegraphics{greybar.ps}}
\end{center}
\caption{The projected mass density in the vicinity of the largest group
in the simulation at $z=0.7$.  The grey scale indicates the projected gas
density (in $h^{-1}M_\odot/(h^{-1}{\rm Mpc})^2$) within a cube of side
$3h^{-1}$Mpc centered on the most bound particle in the group.
Small knots of dense gas, which we identify as sites of galaxy formation, can
be seen clearly.
Like many large groups, this one is part of a large filamentary structure,
which explains the spatial orientation of the distribution of sub-halos.}
\label{fig:greyscale}
\end{figure}

We have used the {\sc Tree/SPH\/} code {\sc Gadget\/}
(Springel, Yoshida \& White~\cite{SprYosWhi}) to run a
$2\times 300^3=54$ million particle simulation of this model in a
periodic box of size $33.5\,h^{-1}$Mpc.  Equal numbers of gas and dark
matter particles were employed, so $m_{\rm dark}=1.0\times 10^8\,h^{-1}M_\odot$
and $m_{\rm gas}=1.5\times 10^7\,h^{-1}M_\odot$.  The gravitational interaction
between particles is softened on small scales using a cubic
spline (e.g.~Hernquist \& Katz~\cite{HerKatz}) and the `Plummer equivalent'
gravitational softening in our simulation was $6\,h^{-1}$kpc, fixed in
comoving coordinates.
The simulation was started at redshift $z=99$ and evolved to $z=0$.
Unfortunately the simulation box is too small to reliably predict 
clustering properties at very low redshifts, as the fundamental mode
becomes increasingly non-linear for $z<1$.
Hence, in this paper we restrict our analysis to redshifts $z\ge 0.5$.

In addition to the gravitational interactions and adiabatic hydrodynamics,
the code follows radiative cooling and heating processes in the presence
of a UV radiation field in essentially the same way as described in
Katz, Weinberg \& Hernquist~(\cite{KatWeiHer}).
We model the UV radiation field using a modified 
Haardt \& Madau~(\cite{HaaMad})
spectrum with reionization occuring 
at $z\approx6$ (see e.g. Dav\'e et al. 1999), and with
an amplitude chosen to reproduce the mean opacity of the Lyman-alpha
forest at $z=3$ (e.g. Rauch et al. 1997).
Star formation (and feedback) is handled using a modification of the
``multi-phase'' model of Yepes et al.~(\cite{YepKatKhoKly}) and
Hultman \& Pharasyn~(\cite{HulPha}).
Each SPH particle is assumed to describe a co-spatial fluid of ambient hot
gas, condensed cold clouds, and stars.  Hydrodynamics is followed for
only the hot gas phase, but the cold gas and stars are subject to gravity,
add inertia, and participate in mass and energy exchange processes with the
ambient gas phase.
The algorithm will be described in more detail in a forthcoming paper.

{}From the simulation outputs we have constructed catalogues of halos and
their sub-halos using the algorithm {\sc Subfind\/} described in detail in
Springel et al.~(\cite{SWTK}).
First, the Friends-of-Friends algorithm (with a linking length of
$0.15\bar{n}^{-1/3}$) is used to define a parent halo catalogue, and then
bound sub-halos within each parent are identified.
Sub-halos are defined as locally overdense, gravitationally bound
structures.
These sub-halos typically consist of cold, dense gas at their center,
surrounded by a halo of dark matter and tenuous hot gas.
The halo may be severely truncated if the sub-halo is not the central galaxy
in the parent halo, but the dense gas that has been able to efficiently cool
and form stars should allow a reliable identification with galaxies in
the real Universe, at least statistically (see Fig.~\ref{fig:greyscale}).

\begin{table}
\begin{center}
\begin{tabular}{ccccc}
$z$ & $N_{\rm sub-halo}$ & $N(>M_{\rm sub})$ & $N(>M_{\rm fof})$ &
  $f_{\rm iso}$ \\
3.00 & 39579 & 7066 &  9742 & 73\% \\
2.00 & 39248 & 9478 & 13339 & 71\% \\
1.75 & 38233 & 9802 & 13752 & 71\% \\
1.50 & 36959 &10129 & 14158 & 70\% \\
1.25 & 35819 &10294 & 14489 & 71\% \\
1.00 & 34422 &10333 & 14514 & 70\% \\
0.70 & 33143 &10165 & 14583 & 70\% \\
0.50 & 32234 & 9958 & 14512 & 70\%
\end{tabular}
\end{center}
\caption{The number of sub-halos found in the simulation as a function of
redshift.  The first column is the total number of sub-halos with more
than 32 particles.  The second column is the number of sub-halos with
$M_{500}>10^{10}\,h^{-1}M_\odot$.  Third, the number of sub-halos which live
in a parent ``FOF'' halo with $M_{500}>10^{10}\,h^{-1}M_\odot$.  The last
column lists the fraction of sub-halos above $10^{10}\,h^{-1}M_\odot$ which
are ``isolated'', i.e.~are the only members of their parent halo.}
\label{tab:nhalo}
\end{table}

We have used a linking length of 0.15 rather than the more canonical 0.2 in
defining the parent halos, because we found for the larger linking length
two neighboring but distinct halos were frequently linked into
one parent halo, with one of them then identified as a sub-halo of the other.
While this problem is not eliminated entirely using a linking length of 0.15,
it is significantly reduced.  Only very close halo pairs or triplets, possibly
in the process of merging, are joined with this linking length.
In some instances the FOF algorithm finds halos which are not bound; these
halos have in general very small particle number and are not included in the
analysis.
Results from the halo finding are shown in Table~\ref{tab:nhalo}.

For each halo or sub-halo we define the ``center'' as the position of the
particle with the minimum potential energy, with the potential calculated
using only the group particles.  This definition usually corresponds very
closely to the most bound and densest particles, and is more robust than the
center of mass.
Many definitions of mass are possible and can differ from each other quite
significantly (e.g.~White~\cite{HaloMass}).
Although counting the particles within the FOF group is the simplest, we have
chosen to follow a more commonly used approach where the mass is defined as
that enclosed within a radius inside of which the mean density is 500 times the
{\it critical\/} density.
An alternative definition could use the background density rather than the
critical density.
The two scale differently with redshift except for cosmologies with critical
matter density.
At the redshifts at which we are working the density contrast with respect to
background scales as $\left[\Omega_{\rm m}+\Omega_\Lambda (1+z)^{-3} \right]$.

\begin{table}
\begin{center}
\begin{tabular}{ccc}
     & \multicolumn{2}{c}{$\log_{10}(M_{500}/M_\odot)$} \\
$z$  & 10-11 & 11-12 \\
3.00 & 76\%  & 27\% \\
2.00 & 76\%  & 16\% \\
1.75 & 76\%  & 19\% \\
1.50 & 76\%  & 17\% \\
1.25 & 77\%  & 18\% \\
1.00 & 77\%  & 19\% \\
0.70 & 77\%  & 17\% \\
0.50 & 77\%  & 19\%
\end{tabular}
\end{center}
\caption{Fraction of sub-halos which are isolated as a function of mass and
redshift.  There are no halos in the simulation with
$M_{500}>10^{12}\,h^{-1}M_\odot$ which are isolated for $z\ge 0.5$!}
\label{tab:isofrac}
\end{table}

\begin{figure}
\begin{center}
\resizebox{3.5in}{!}{\includegraphics{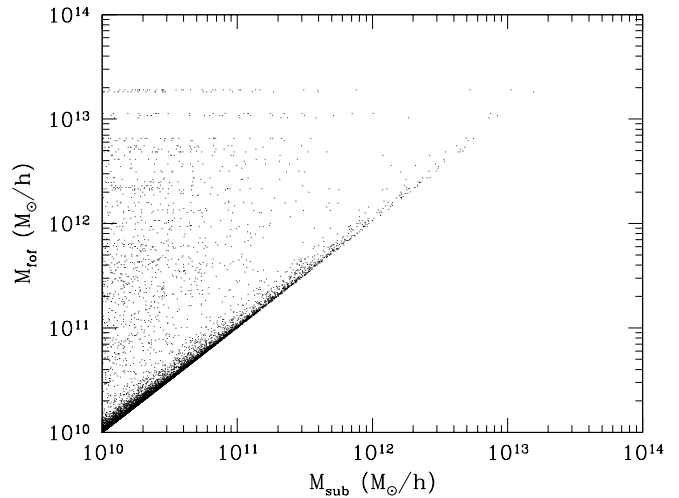}}
\end{center}
\caption{A scatter plot of $M_{500}$ for the sub-halos and their parent
FOF halos from the simulation at $z=1$.}
\label{fig:massmass}
\end{figure}

In Fig.~\ref{fig:massmass} we show a scatter plot of the masses of the
sub-halos and their parent halos at $z=1$.
Most of the halos are isolated and have parent halo masses fractionally larger
than the sub-halo masses.  As the parent halos become more massive there is
less chance that it hosts only one galaxy.
In our first pass through we found that there are a very small number of
systems where the sub-halo mass exceeds the parent halo mass.  This can arise
in situations where several halos are merging, with the bridging material
being at low density.  If the most bound particle in the group lies at the
center of a less massive sub-halo our definition of mass, $M_{500}$, returns
the mass of this sub-halo as the ``parent'' mass.
In such cases we replace the ``parent'' mass with the sum of the sub-halo
masses to more accurately reflect the total mass of the system.

\begin{table}
\begin{center}
\begin{tabular}{cc}
$z$ & $\langle z_{\rm form}\rangle$ \\
3.00 & 4.8 \\
2.00 & 3.3 \\
1.75 & 3.0 \\
1.50 & 2.7 \\
1.25 & 2.4 \\
1.00 & 2.2 \\
0.90 & 2.1 \\
0.80 & 2.0 \\
0.70 & 1.9 \\
0.60 & 1.8 \\
0.50 & 1.7
\end{tabular}
\end{center}
\caption{The (mass weighted) mean redshift of star formation as a function
of redshift.}
\label{tab:zform}
\end{table}

\begin{figure}
\begin{center}
\resizebox{3.5in}{!}{\includegraphics{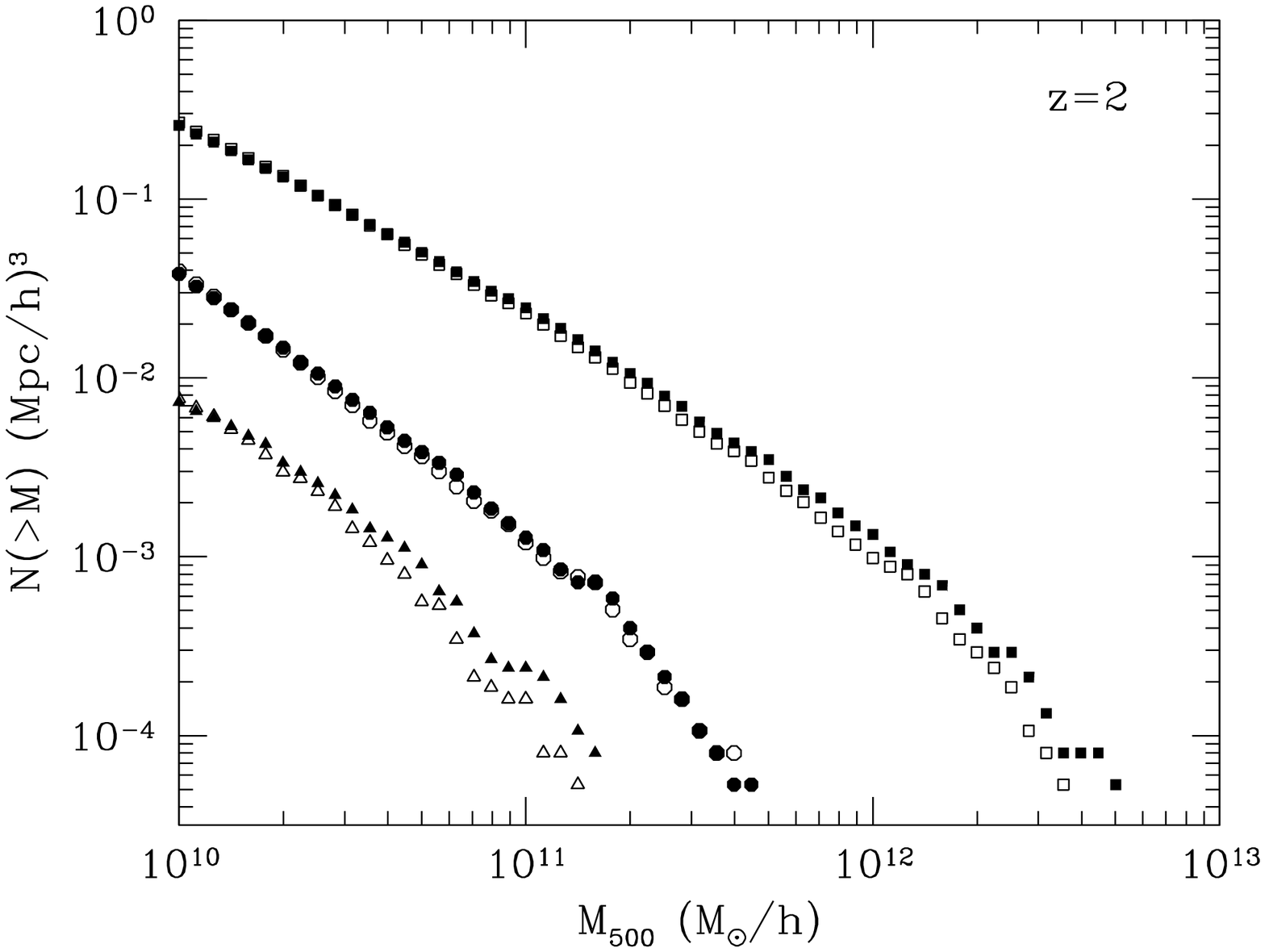}}
\resizebox{3.5in}{!}{\includegraphics{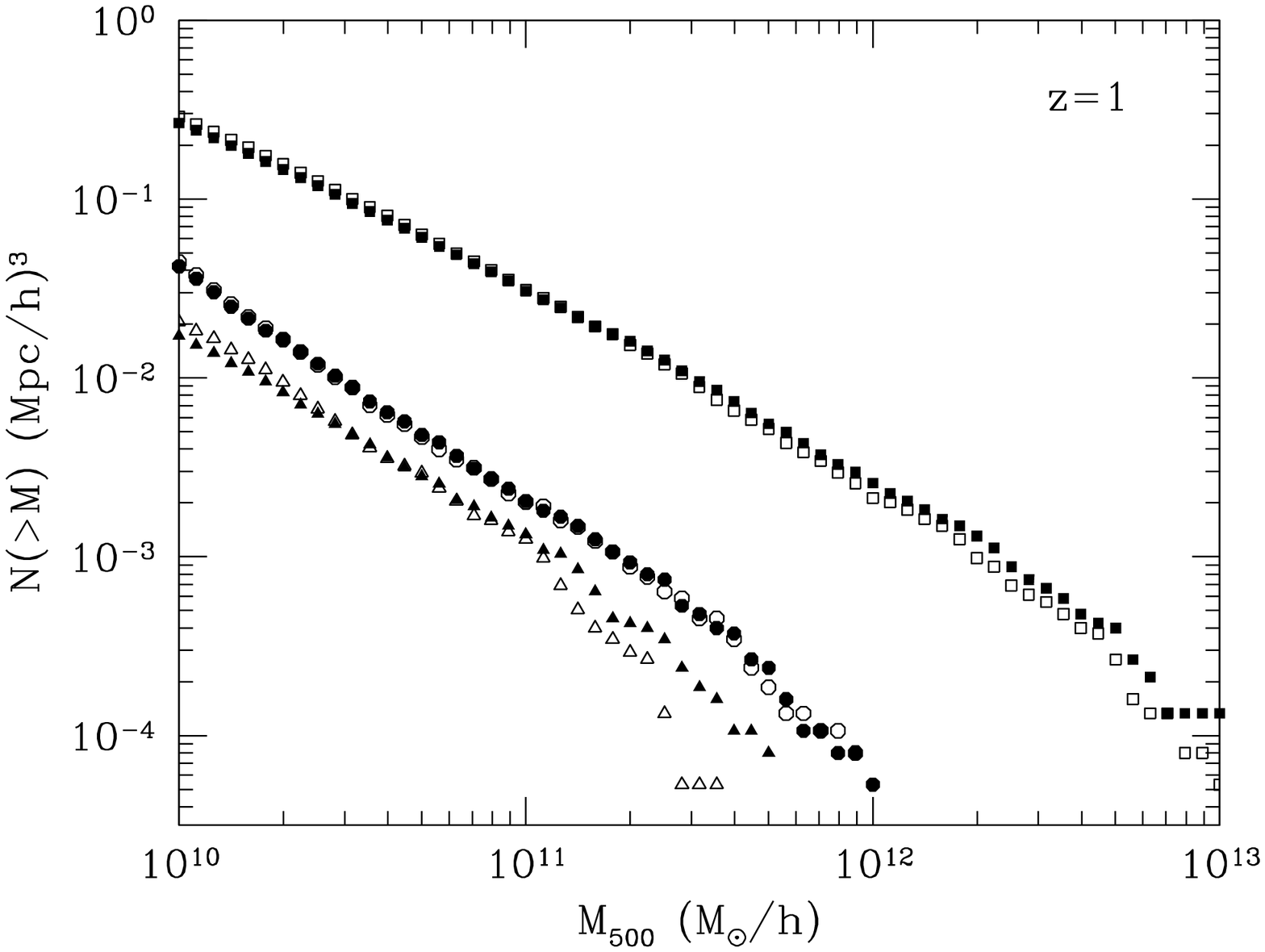}}
\resizebox{3.5in}{!}{\includegraphics{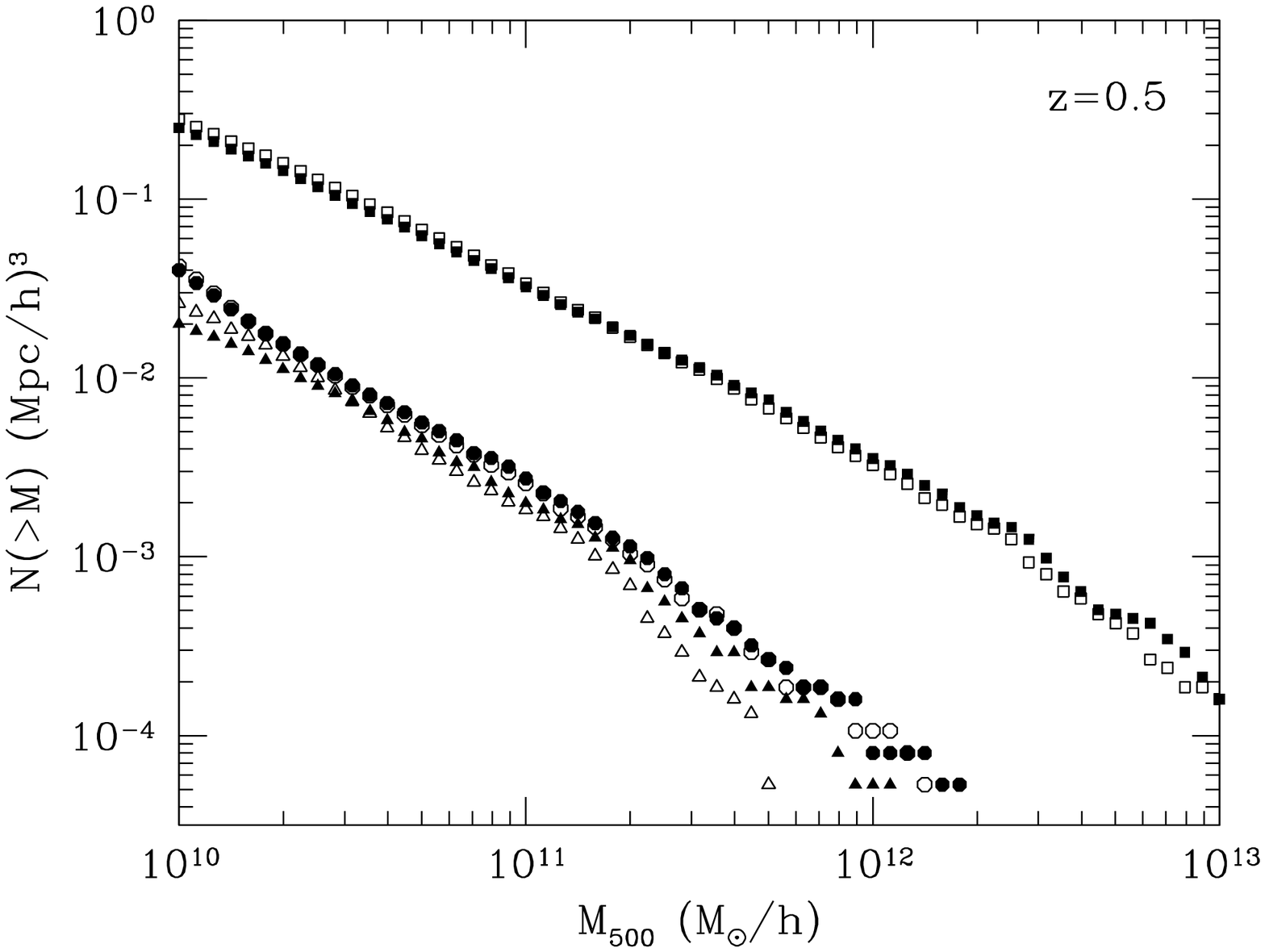}}
\end{center}
\caption{The mass function for the simulation at $z=2$ (top), 1 (middle)
and $0.5$ (bottom).
Solid symbols are for the ``parent'' halos found with FOF with a linking
length of 0.15.  Open symbols count halos identified by the sub-halo finder,
removing unbound particles.  We show the mass function as a function of total
mass (squares), gas mass (circles) and stellar mass (triangles).  In all cases
the mass quoted is $M_{500}$ in $h^{-1}\,M_\odot$ (see text).}
\label{fig:nofm_z}
\end{figure}

Ideally, we would identify the sub-halos as galaxies by some observational
property, such as luminosity or color.
Unfortunately, it is difficult to reliably compute such properties from this
simulation, because the outputs are too infrequent to graft on population
synthesis codes.
However we do give, in addition to the total mass of a sub-halo, results as
a function of the stellar mass within $r_{500}$ of the sub-halo center.
We do this under the assumption that the near infrared magnitude of a galaxy
should roughly track its stellar mass (e.g.~Cole et al.~\cite{2dF-Kband}).
The (mass weighted) mean age of the stars at each output is given in
Table~\ref{tab:zform} to help with this conversion.
Finally, to help in matching our halos to observed galaxies we give the number
density of our halos, as a function of $z$, in Fig.~\ref{fig:nofm_z}.
As a first approximation, and in the absence of further information, one
can match halos of a given mass to objects of an equivalent space density.

\section{The clustering of sub-halos} \label{sec:clustering}

\begin{figure}
\begin{center}
\resizebox{3.5in}{!}{\includegraphics{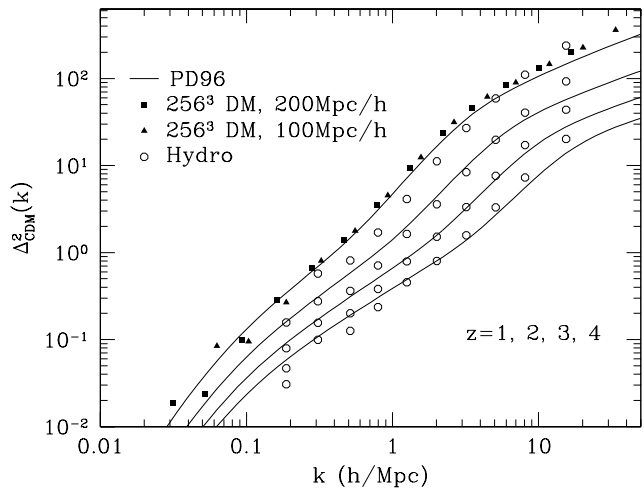}}
\end{center}
\caption{The dimensionless dark matter power spectrum as a function of
redshift.  The solid lines indicate the expected non-linear power spectrum
for this model, as estimated using the formalism of
Peacock \& Dodds~(\protect\cite{PD96}) at $z=1$, 2, 3 and 4 (top to bottom).
The solid symbols are the power spectra at $z=1$ from two $256^3$ DM only
simulations in boxes of size $200h^{-1}$Mpc (squares) and $100h^{-1}$Mpc
(triangles) which provide and estimates of the uncertainty 
in the Peacock \& Dodds formalism.
The open circles indicate the power spectra from the simulation
in this work, the $33.5h^{-1}$Mpc hydro simulation, at $z=1$, 2, 3 and 4
(top to bottom).}
\label{fig:pk_z}
\end{figure}

We are interested in the clustering of galaxies and dark matter, and its
evolution with time, within this simulation.
We show in Fig.~\ref{fig:pk_z} the dark matter\footnote{The power spectrum
of the gas mass follows the dark matter power spectrum to
$k\ga 10h{\rm Mpc}^{-1}$.  The stellar particles are more strongly clustered
as expected.} power spectra at $z=4$, 3, 2 and 1.
To compute these power spectra we estimated the dark matter density field from
the particles using NGP assignment 
(Hockney \& Eastwood~\cite{HocEas}) onto a $256^3$ grid.
The density field was then Fourier Transformed to obtain $\delta_k$.
The power spectrum was obtained by binning $\left| \delta_k \right|^2$ in
shells in $\vec{k}$-space, corrected for the binning, the mass assignment onto
the grid and shot-noise (see e.g.~Peacock \& Dodds~\cite{PD96} for a discussion
of some of these issues).
To extend the calculation to higher $k$ we repeated the process several times,
rescaling the particle separations by increasing factors and remapping the
distribution into the periodic volume each time
(Peacock, private communication; Jenkins et al.~\cite{Jen98}).
This provides us with an estimate of $P(k)$ limited only by the resolution of
the simulation, and not by the size of the Fourier transform grid.

\begin{figure}
\begin{center}
\resizebox{3.5in}{!}{\includegraphics{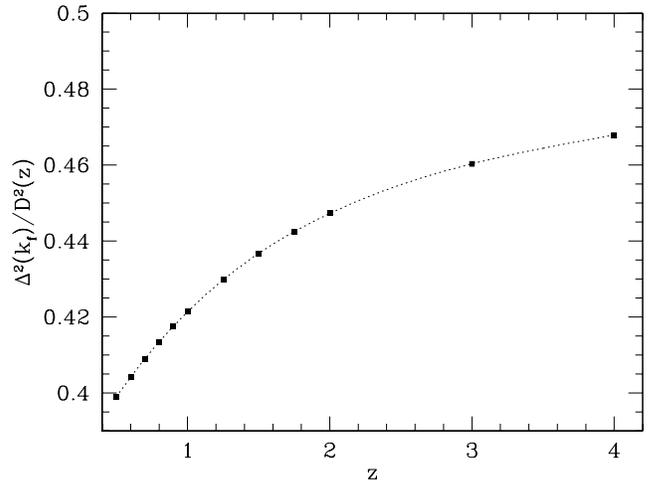}}
\end{center}
\caption{The power in the fundamental mode vs linear theory.  We plot the
ratio of $\Delta^2(k_f)$ to the linear growth factor, $D(z)$.  This should
be constant as long as the box is a fair sample of the Universe.  The power
begins to fall below the linear theory prediction at late times because our
periodic box is missing long-wavelength power on scales larger than the
fundamental mode, which are not available to be coupled into the evolution
as $k_f$ goes non-linear.}
\label{fig:kf}
\end{figure}

Compared to semi-analytic estimates of the expected power
(Peacock \& Dodds~\cite{PD96})  we find that this box has a slight shortfall.
To check whether the semi-analytic model correctly estimates the dark matter
power spectrum for this model we have run two additional dark matter only
simulations, each using $256^3$ particles.  The first simulation is in a
$200h^{-1}$Mpc box and the second in a $100h^{-1}$Mpc box.  The power spectra
computed from the $z=1$ outputs of these runs in the same manner as above are
also shown in Fig.~\ref{fig:pk_z}.
We see that the fitting function has slightly less power than the simulation
at small scales, as seen also in the simulations of
Jain, Mo \& White~(\cite{JaiMoWhi}).
The power spectrum in our $33.5h^{-1}$Mpc box is low on all scales at the
redshifts of interest, although the shape is approximately correct.
Fluctuations in the amplitude, but not the shape, of the power spectrum due
to finite volume effects are well known (e.g.~Meiksin \& White~\cite{MeiWhi}).
Part of the shortfall in power is due to the particular random phases chosen
in the initial conditions, the remainder is due to the fact that the box is
becoming less and less of a fair sample of the Universe as time evolves and
the non-linear scale increases.
We show the latter effect by plotting, in Fig.~\ref{fig:kf}, the growth of
the fundamental mode, $\Delta^2(k_f)$, compared with linear theory.
As long as the box remains a fair sample of the Universe the ratio
$\Delta^2(k_f)/D^2(z)$, where $D(z)$ is the linear growth factor, should
remain constant.

In earlier work (White et al.~\cite{WhiHerSpr}) we showed that the number of
``galaxies'' per dark matter halo scaled approximately as a power-law in the
parent halo mass with an index less than 1 (typically 0.7-0.8).
This behavior is precisely what is needed to explain the observed clustering
of galaxies (e.g.~Scoccimarro et al.~\cite{SSHJ}) and is also seen in
semi-analytic models (e.g.~Seljak~\cite{Sel}).
The spatial distribution of sub-halos in the simulation is consistent with
the assumption that every halo hosts a galaxy which resides at its center
and any extra satellite galaxies trace the dark matter distribution.
It remains an open question observationally whether the galaxy density traces
the mass density in halos (e.g.~Adami et al.~\cite{AMUS}).

We performed a counts-in-cells analysis of both the mass and galaxy number
density fields as a function of redshift.  First, we divided the cubical box
up into a grid of cells with $N_c=32$, 64, 128 or 256 cells on a side.  We
assigned each particle or galaxy to the appropriate cubical cell weighting
the particles by their mass and the galaxies equally.  For the galaxies, we
treated them as points at the position of their potential minima.  This was
done 8 times with the grid shifted to a random position each time.  Then, we
calculated the variance of mass and galaxy fluctuations, from which we can
define $b\equiv\sigma_{\rm gg}/\sigma_{\rm mm}$, the covariance
$b'\equiv \sigma^2_{\rm gm}/\sigma_{\rm mm}^2$ and the cross-correlation
coefficient $r\equiv \sigma^2_{\rm gm}/\sigma_{\rm gg}\sigma_{\rm mm}$.
Our results are shown in Table~\ref{tab:cic}.  We see a clear tendency for the
bias to be ``stochastic'', with decreasing correlation between the galaxies and
the total mass as we probe smaller length scales.

\begin{table}
\begin{center}
\begin{tabular}{ccccc}
$N_c$ & $\sigma_{\rm mm}$ & $\sigma_{\rm gg}$ & $r$ \\
  16    &  1.87  &  1.59 & 0.83 \\
  32    &  3.56  &  3.02 & 0.73 \\
  64    &  6.59  &  6.45 & 0.61 \\
 128    &  11.2  &  15.7 & 0.51 \\
 256    &  18.0  &  41.6 & 0.46
\end{tabular}
\end{center}
\caption{Counts-in-cells of the mass and galaxy number density fields at
$z=1$.  The first column is the number of cells in each dimension of the box,
the second column gives the rms fluctuation in the mass (in units of the
mean), the third the rms fluctuation in the galaxy number density (using only
galaxies with $M_{500}>10^{10}\,h^{-1}M_\odot$) and the final column gives
the cross correlation coefficient.  We have stopped at $N_c=256$ since to
implement finer grids we would need to use the extent of the sub-halos rather
than just their positions.}
\label{tab:cic}
\end{table}

\section{Correlation function} \label{sec:corrfn}

We identify the center of a ``galaxy'' as the position of the minimum of
the potential of a bound sub-halo found with {\sc Subfind}.
Around each such center we calculate the probability, in excess of random,
of having mass $M$ within a spherical shell of radius $r$ and width $dr$.
Stellar mass, cold and hot gas mass and dark matter are all included in
this accounting.  We present our results in terms of the correlation
function which we define as
\begin{equation}
  \xi_{\rm gm} \equiv {M(r;dr)\over \langle M(r;dr)\rangle} -1 ,
\end{equation}
where $M(r;dr)$ is the mass contained within the shell between radius $r$
and $r+dr$ and $\langle\cdots\rangle$ indicates an average quantity.
To safely avoid numerical resolution effects we shall consider galaxies only
above $10^{10}\,h^{-1}M_\odot$.

\begin{figure}
\begin{center}
\resizebox{3.5in}{!}{\includegraphics{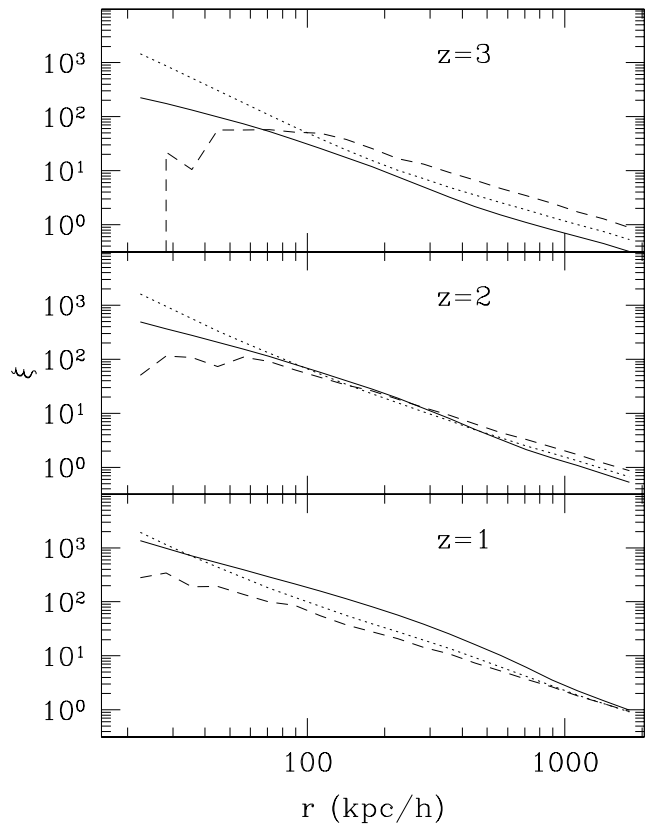}}
\end{center}
\caption{The 2-point correlation functions of the mass (solid lines),
galaxies with $M_{500}>10^{10}h^{-1}M_\odot$ (dashed) and the galaxy-mass
cross correlation function (dotted) at $z=1$, 2 and 3.}
\label{fig:xi_z}
\end{figure}

\begin{figure}
\begin{center}
\resizebox{3.5in}{!}{\includegraphics{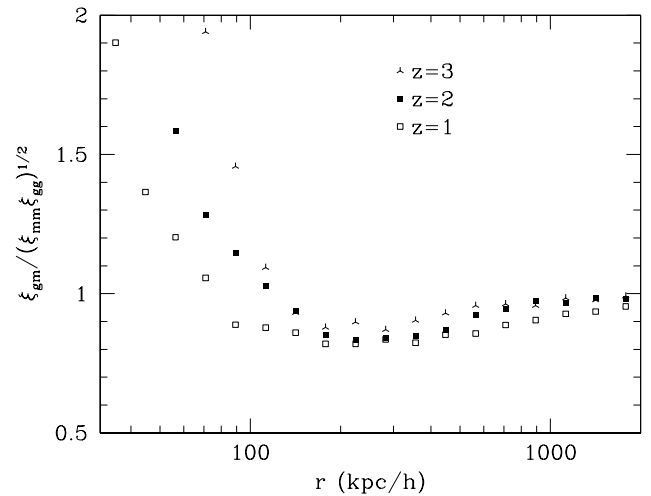}}
\end{center}
\caption{The cross correlation coefficient for the parameters in
Fig.~\protect\ref{fig:xi_z}.  The signal is noisy for high redshift and
small distance, due to the finite number of ``galaxies'' and mass resolution
limitations in the simulation.}
\label{fig:r}
\end{figure}

We show in Fig.~\ref{fig:xi_z} the correlation function of the mass,
$\xi_{\rm mm}(r)$, the galaxies above $10^{10}h^{-1}M_\odot$,
$\xi_{\rm gg}(r)$, and the galaxy-mass cross correlation, $\xi_{\rm gm}(r)$,
at $z=1$, 2 and 3.
The correlation functions are well described by power laws (with a slowly
changing slope) and on scales above a few$\times 100h^{-1}$kpc the galaxy-mass
cross correlation function is just the geometric mean of the mass-mass and
galaxy-galaxy autocorrelation functions (see Fig.~\ref{fig:r}).
While the mass correlation function grows steadily with time the evolution of
the galaxy correlation function is more complicated.
We give the correlation length, $r_0$, defined as $\xi(r_0)=1$, vs.~redshift
in Table~\ref{tab:r0}.  Note that on length scales approaching $1h^{-1}$Mpc
our box is missing power (as described above) and so these correlation lengths
are biased low.
These results are broadly consistent with the clustering properties of
``galaxies'' in simulations reported by Katz et al.~(\cite{KHW92,KHW99})
once differences in mass resolution, box size and galaxy identification
are taken into account.

Finally, we make a distinction between two types of galaxies, those that are
the sole resident of a dark matter halo (``isolated galaxies'') and those
which are members of a larger halo containing several sub-halos
(see Tables~\ref{tab:nhalo}, \ref{tab:isofrac}).  As expected the more
massive galaxies reside in the more massive parent halos which host more
than one sub-halo and so the isolated fraction decreases with mass.  Most
of the lower mass sub-halos are isolated.
As we shall see below, we expect the traditional interpretation of
galaxy-galaxy lensing to be more correct for the isolated galaxies, which in
our case means those of low mass.

\begin{table}
\begin{center}
\begin{tabular}{ccccc}
	& \multicolumn{4}{c}{$r_0$} \\
$z$	& fit 	& mm	& gm	& gg \\ \hline
3.00	&  900	&  765	& 1130	& 1700 \\
2.00	& 1500	& 1200	& 1350	& 1600 \\
1.75	& 1750	& 1300	& 1400	& 1600 \\
1.50	& 2000	& 1400	& 1500	& 1600 \\
1.25	& 2400	& 1600	& 1600	& 1650 \\
1.00	& 2800	& 1800	& 1700	& 1700 \\
0.70	& 3350	& 2200	& 2000	& 1900 \\
0.50	& 3800	& 2550	& 2300	& 2000
\end{tabular}
\end{center}
\caption{The correlation length $r_0$ defined by, $\xi(r_0)=1$, in $h^{-1}$kpc,
as a function of redshift.  The first column (fit) gives the dark matter
auto-correlation length computed from the fitting function of
Peacock \& Dodds~(\protect\cite{PD96}), the other three columns give the
lengths computed from the simulation.
Note the shortfall in power, discussed in the text.
The mass correlation function $\xi_{\rm mm}$ increases monotonically with
time while the galaxy auto-correlation function (for galaxies with
$M_{500}>10^{10}h^{-1}M_\odot$), $\xi_{\rm gg}$, and galaxy-mass
cross-correlation, $\xi_{\rm gm}$, show more complicated evolution.}
\label{tab:r0}
\end{table}

\section{Galaxy-galaxy lensing} \label{sec:gglensing}

\begin{figure}
\begin{center}
\resizebox{3.5in}{!}{\includegraphics{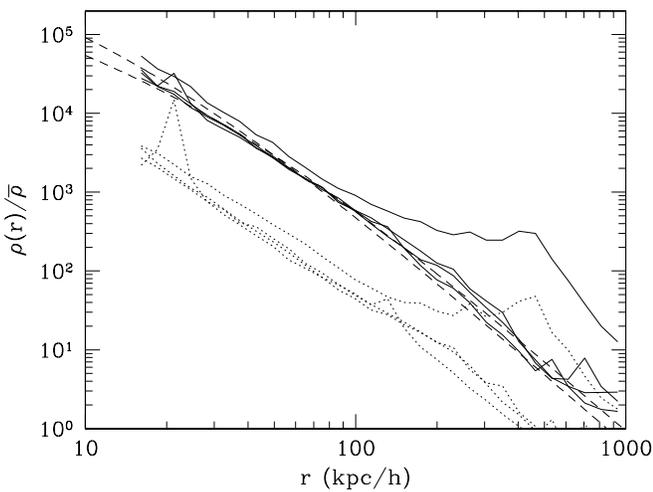}}
\end{center}
\caption{The spherically averaged mass profile, scaled to the background
density $\bar{\rho}=\Omega_{\rm m}\rho_{\rm crit}$, for the 4 ``galaxies''
with $M_{500}$ within 5\% of $10^{12}h^{-1}M_\odot$, extracted from the $z=1$
output.  The solid lines include all of the mass, the dotted lines only the
baryonic (gas$+$stellar) mass.  The dashed lines show NFW profiles
with $c=5$ and 10 (see Eq.~\protect\ref{eqn:nfw}).
The secondary ``peaks'' in the profiles are due to neighboring halos -- as
discussed in the text none of these ``galaxies'' are isolated though all are
centrally located in the parent halo.}
\label{fig:profile}
\end{figure}

The observational study of galaxy-galaxy lensing has a long history
(Tyson et al.~\cite{TVJM})
although only recently have definitive detections been made by several groups
(Brainerd et al.~\cite{BraBlaSma};
 Dell'Antonio \& Tyson~\cite{DelTys};
 Griffiths et al.~\cite{GCIR};
 Hudson et al.~\cite{HGDK};
 Natarajan et al.~\cite{NKSE};
 Wilson et al.~\cite{WKLC};
 Fischer et al.~\cite{SDSS}).
Such studies have traditionally been interpreted as constraints on extended
dark matter halos around ``typical'' galaxies (see Fig.~\ref{fig:profile}).
A more modern interpretation, within the context of large-scale structure,
is as a projection of the 3D galaxy-mass correlation function
(Kaiser~\cite{Kai92}).  As with any such projection, the effect of material
along the line-of-sight but not physically associated with the object in 
question can be a serious one
(see e.g.~Metzler, White \& Loken~\cite{MetWhiLok}).
We shall not address this issue here, focusing instead on the underlying 3D
correlation function itself and its interpretation.

Similar work on galaxy-galaxy lensing has already been presented by
Guzik \& Seljak~(\cite{GuzSel}) using semi-analytic models of galaxy formation.
By comparison with the semi-analytic models, our simulations have better
spatial and mass resolution and include far more physics.  Unfortunately
limitations on computer resources have forced us to simulate a relatively
small volume of space.
This both reduces the size of our samples for statistical purposes and limits
the minimum redshift to which we can accurately follow the development of
large-scale structure.
It is thus encouraging that our results are in good agreement with theirs in
many respects.

\begin{figure}
\begin{center}
\resizebox{3.5in}{!}{\includegraphics{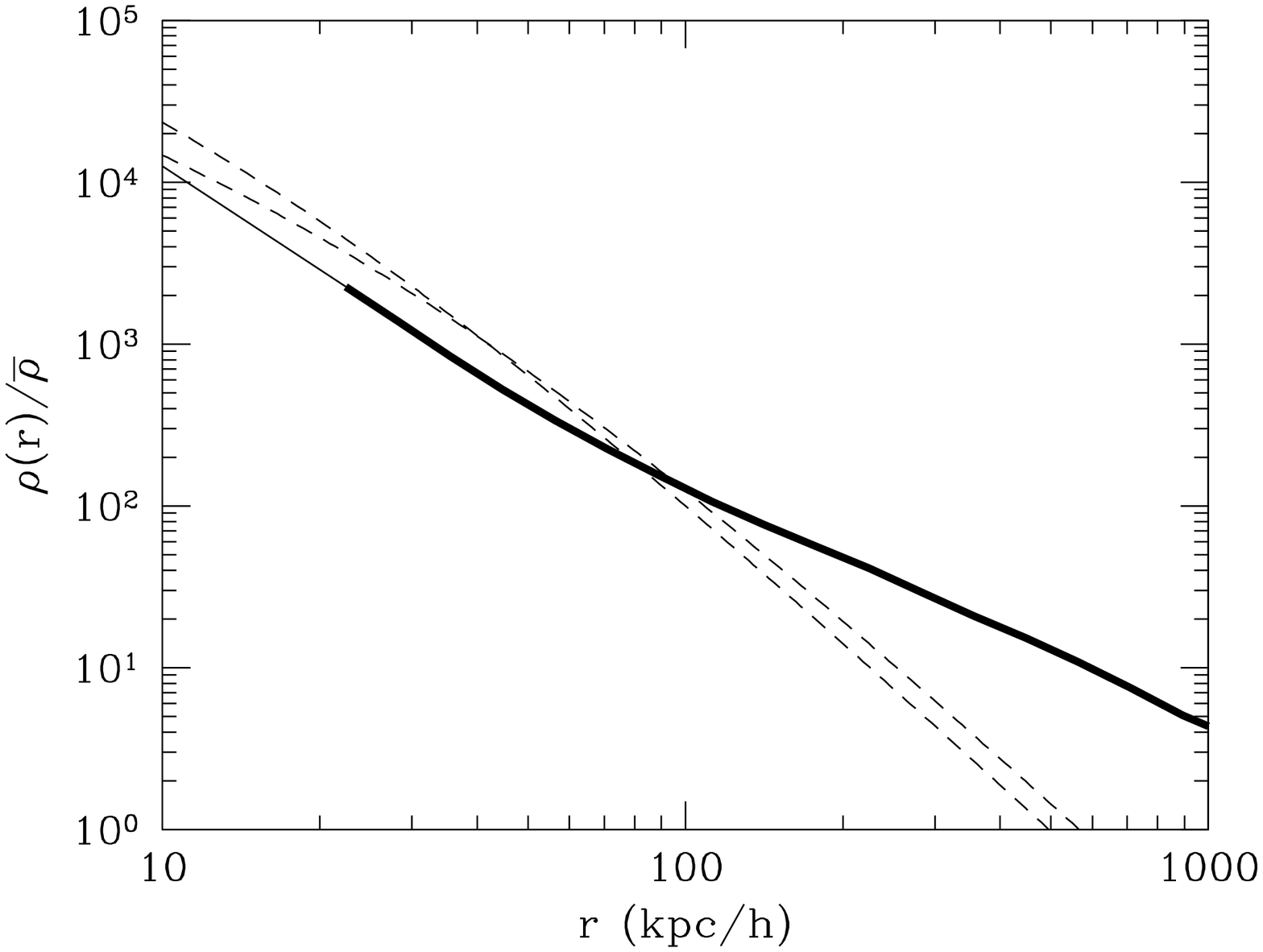}}
\resizebox{3.5in}{!}{\includegraphics{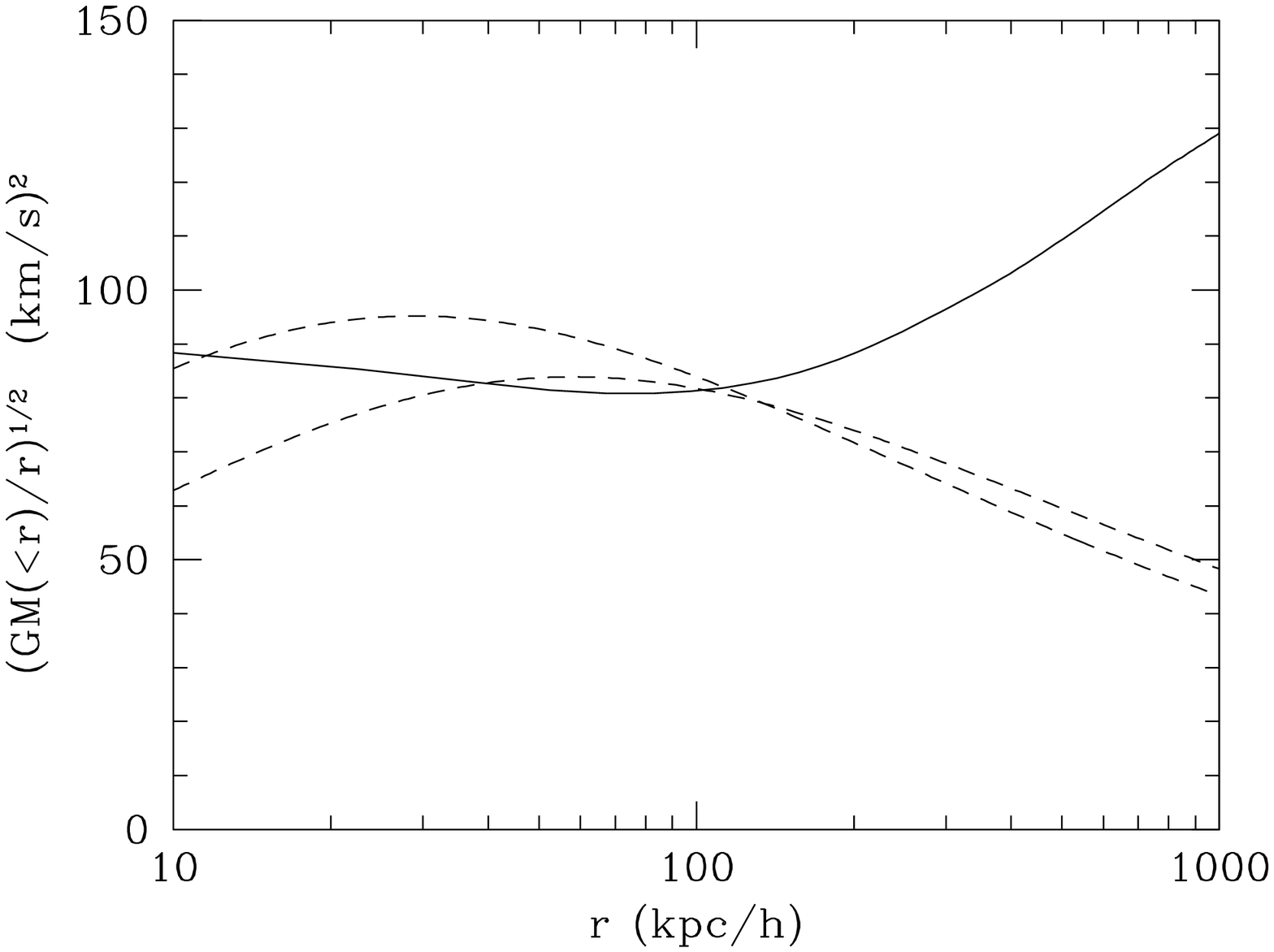}}
\end{center}
\caption{(top) The profile obtained from the simulation at $z=0.5$ if we
interpret $\xi_{\rm gm}(r)$ as a density profile through
$\rho_{\rm eff}(r)\equiv \bar{\rho}\left(1+\xi_{\rm gm}(r)\right)$.
The bold curve indicates the region where we have explcitly computed
$\xi_{\rm gm}$ from the simulation, the rest is power-law extrapolation.
The dashed lines are NFW profiles with
$M_{\rm vir}=2\times 10^{11}\,h^{-1}M_\odot$ and $c=5$ and 10 (see text).
(bottom) The rotation curve from this profile, assuming $v_c=\sqrt{GM/r}$,
and for the two NFW profiles.}
\label{fig:makeprofile}
\end{figure}

As a first step we therefore ask how $\xi_{\rm gm}(r)$ relates to the
profile of an ``average'' galaxy in our simulation?
As shown in Fig.~\ref{fig:profile}, the spherically averaged profiles
of our halos are reasonably well fit by the NFW profile
(Navarro et al.~\cite{NFW}):
\begin{equation}
  \rho(r) = {\rho_0\over x(1+x)^2} \qquad ,
\label{eqn:nfw}
\end{equation}
where $x=r/r_s$ is the radius scaled in units of a characteristic radius $r_s$.
The central density is fixed by specifying the halo mass.
While we have worked throughout in terms of $M_{500}$, the mass of the halo is
typically defined to be the virial mass
$M=(4\pi/3) \delta_{\rm vir}\rho_{\rm crit}r_{\rm vir}^3$ where
we take the virial radius, $r_{\rm vir}$, as the radius within which the
mean density enclosed is $\delta_{\rm vir}$ times the critical density.
Since our cosmology has $\Omega_{\rm m}\ne 1$, the top-hat model prediction
for $\delta_{\rm vir}$ is redshift dependent, taking the value
$\delta_{\rm vir}\simeq 100$ at $z=0$, $\delta_{\rm vir}\simeq 140$ at
$z=0.5$ and $\delta_{\rm vir}\to 18\pi^2$ as $z\to\infty$.
The parameter $c\equiv r_{\rm vir}/r_s$ measures the degree of central
concentration of this mass.

To estimate a profile from the cross correlation we define
$\rho_{\rm eff}(r)\equiv \bar{\rho}\left(1+\xi_{\rm gm}(r)\right)$,
using a power-law extrapolation to extend the range for both large and small
radius beyond what we have computed explicitly from the simulation.
The results are shown in Fig.~\ref{fig:makeprofile} along with two NFW
profiles of virial mass $M_{140}=1.8\times 10^{11}\,h^{-1}M_\odot$, which is
the ``average'' mass of halos in the simulation at $z=0.5$.

The concept of an ``average'' mass is somewhat nebulous.  Simply averaging
$M_{500}$ for the halos identified at $z=0.5$ gives
$\langle M_{500}\rangle=9\times 10^{10}h^{-1}M_\odot$.
Relating this to a virial mass is complicated by the fact that the
concentration, $c$, and thus the ration $M_{\rm vir}/M_{500}$,
varies with mass.  An alternative route is to use the
Jenkins et al.~(\cite{JFWCCEY}) fit to the mass function to compute
\begin{equation}
  \langle M\rangle \equiv { \int_{M_{\rm cut}}^\infty M dn\over
                            \int_{M_{\rm cut}}^\infty \ dn}
\end{equation}
above $M_{\rm cut}=1.4\times 10^{10}\,h^{-1}M_\odot$.  The limiting mass is
obtained by converting $M_{500}=10^{10}\,h^{-1}M_\odot$ to $M_{140}$ for an
NFW profile with $c=10$.  [We find that for $z>0.5$ the virial mass and
$M_{200}$ differ by less than 10\%, so one could alternatively use $M_{200}$
throughout.]  We employ this conversion because the virial mass is closer to
the ``mass'' definition used by Jenkins et al.~(\cite{JFWCCEY}) than the
$M_{500}$ values preferred in this study.
While several steps are involved here, White~(\cite{HaloMass}) has shown that
using an NFW profile to convert between such mass definitions works very well.
Varying the value of the concentration parameter in our conversion from
$M_{500}$ to $M_{140}$ makes little difference to our average mass.
For a concentration of $5$ the mass ratio is $M_{140}=1.6M_{500}$ and
$M_{\rm cut}=2\times 10^{11}\,h^{-1}M_\odot$.

As we can see the NFW profiles, while providing a good fit to individual
``galaxies'' within the simulation (e.g.~Fig.~\ref{fig:profile}) do not
provide a very good fit to the profile $\rho_{\rm eff}(r)$.
Also, using $\rho_{\rm eff}(r)$ we would estimate that the virial mass of
an ``average'' galaxy is $M_{102}=1.0\times 10^{11}\,h^{-1}M_\odot$
($M_{500}=5.2\times 10^{10}\,h^{-1}M_\odot$, c.f.~$9\times 10^{10}$ above),
a factor of two lower than the average computed above.
Finally we note that the associated rotation curve for our ``average'' galaxy
is much flatter than the rotation curves for the individual galaxies making up
the average.  This is a consequence of the varying virial radii and the
weighting by the mass function and is {\it not\/} reflecting the distribution
of matter in the individual halos.

These results suggest that galaxy-galaxy lensing, wherein one increases the
signal-to-noise by stacking many galaxies, is not in fact measuring the
profile of an ``average'' galaxy in the usual sense of these words.  The
main problem is that galaxies come in a wide range of masses and sizes,
and live in a range of environments so the naive averaging loses its
significance.  If it were possible to restrict the galaxies going into the
average then a more faithful representation could be obtained (although
projection effects could still be a significant source of error), but we
have no a priori way of knowing the galaxy mass.
These issues should be borne in mind when attempting to use a weak lensing
analysis to estimate halo profiles or mass-to-light ratios from galaxy-galaxy
lensing.

There are two effects which muddy the waters.  First, not all galaxies are the
sole members of their dark matter halos.
Even for those galaxies which are ``isolated'' what one measures is the
integral over the mass function of halo profiles.  Since more massive halos
are in general larger, there is no sense in which one measures an ``average''
profile.
For example, Seljak~(\cite{Sel}) has suggested that much of the large-$r$
signal in this case comes from the larger halos rather than the asymptotic
behavior of the smaller halos.

\begin{figure}
\begin{center}
\resizebox{3.5in}{!}{\includegraphics{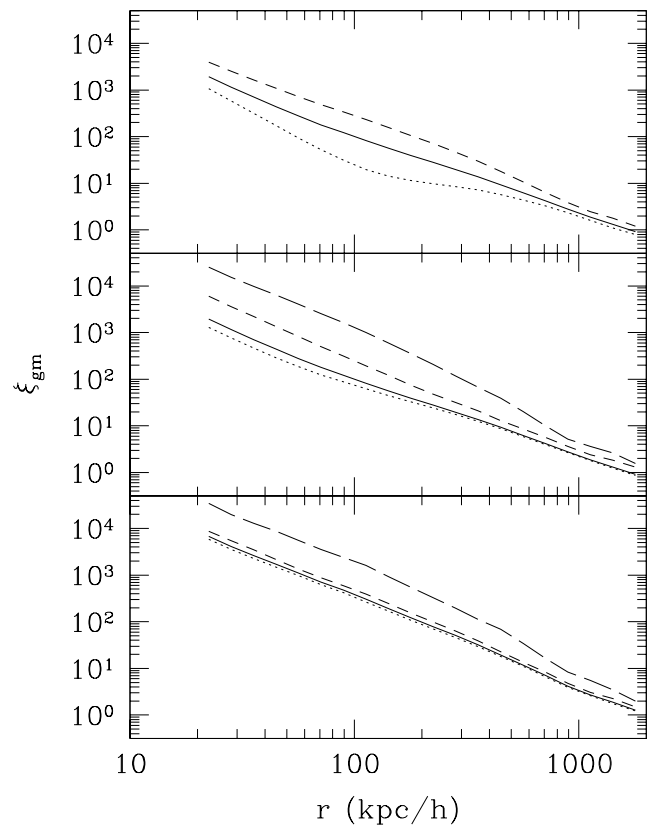}}
\end{center}
\caption{The galaxy-mass cross correlation, $\xi_{\rm gm}(r)$ at $z=1$.
In each panel the solid line is the cross correlation including all massive
galaxies.  In the top panel, we show separately the contributions from
``isolated'' galaxies (dotted lines) and from galaxies sharing halos
(dashed lines).  In the middle panel we show the cross correlation around
galaxies with total masses
$10^{10}h^{-1}\,M_\odot\le M_{500}< 10^{11}h^{-1}\,M_\odot$ (dotted),
$10^{11}h^{-1}\,M_\odot\le M_{500}< 10^{12}h^{-1}\,M_\odot$ (short dashed),
$10^{12}h^{-1}\,M_\odot\le M_{500}< 10^{13}h^{-1}\,M_\odot$ (long  dashed).
The more massive sub-halos are rare, leading to a noiser signal.
In the lower panel we show the cross correlation around galaxies with stellar
masses
$10^{ 9}h^{-1}\,M_\odot\le M_{500}< 10^{10}h^{-1}\,M_\odot$ (dotted),
$10^{10}h^{-1}\,M_\odot\le M_{500}< 10^{11}h^{-1}\,M_\odot$ (short dashed),
$10^{11}h^{-1}\,M_\odot\le M_{500}< 10^{12}h^{-1}\,M_\odot$ (long  dashed).
The situation at higher and lower redshifts is very similar to that shown
here.}
\label{fig:xigm}
\end{figure}

In Fig.~\ref{fig:xigm} we show that, although the ``isolated'' galaxies
are the majority of the number counts (Table~\ref{tab:nhalo}, see also
White et al.~\cite{WhiHerSpr}), the signal for $\xi_{\rm gm}(r)$ is
intermediate between that for isolated and non-isolated groups.
The two curves begin to merge at very large radius where the density is
becoming close to the cosmic mean.
This suggests that at larger radius ($r\sim 1h^{-1}$Mpc) we are probing not the
galactic mass profile but the general distribution of matter and the isolated
and non-isolated galaxies have similar large-$r$ profiles.
Further,the fact that $\xi_{\rm gm}$ differs from the isolated halo case
at smaller $r$ indicates that even there one is not measuring the profile of
the galactic halo.
Galaxies which are part embedded in larger halos contribute disproportionately
to the signal from $10h^{-1}$kpc out to several $100h^{-1}$kpc.
It is this ``intermediate'' range of radii where the signal
is dominated by larger halos, as Seljak~(\cite{Sel}) suggests.
In either case, when one measures $\xi_{\rm gm}(r)$ one is not simply measuring
the mass profile around a galaxy located at the center of its dark matter halo.

Guzik \& Seljak~(\cite{GuzSel}) have suggested that isolated galaxies should
show stronger cross-correlation at small-$r$ and fixed mass, since they are
all centrally located within their halos.  At fixed galaxy (i.e.~sub-halo)
mass the opposite effect is true, since the non-isolated galaxies tend to
live in more massive halos.
However, if we hold the mass of the parent halo rather than that of the
galaxy fixed we can correct for this effect.
We find that if we do this then the central galaxies are often in denser
environments, in the sense that $\xi_{\rm gm}$ is larger at very small
$r\sim 10\,h^{-1}$kpc.  However by $r\sim 100\,h^{-1}$kpc the opposite effect
is true and the non-isolated galaxies have the stronger cross correlation.
This persists until $r\sim 1\,h^{-1}$Mpc where the cross correlations for
isolated and non-isolated galaxies are equal.
We also note that the density around the peak is determined primarily by the
mass and formation time of the sub-halo, not on whether it is part of a larger
structure.  Second, many of our systems contain distinct sub-halos, in
which however the ``bridging'' material is at quite high density (indicating
that it really is part of a larger structure and not an artifact of our group
finding technique).
In some sense all of these sub-halos are ``central'' and have a
similar mass profile even though they belong to a larger halo.
This shows that the idealization of spherical parent halos is a poor
approximation to the groups we see in the simulation.

To understand the effect of the integration over mass we show the signal
broken up in different ways in Fig.~\ref{fig:xigm}.
In the middle panel of Fig.~\ref{fig:xigm} we break things down by the mass
of the galaxy itself.
Since there is a trend for more massive galaxies to be larger and to live in
more massive parent halos, their cross-correlation is larger.  Again, even
though the vast majority of galaxies (by number) have low mass, the signal
comes from a range of masses.
We also break the signal down by the galaxy {\it stellar\/} mass, under the
assumption that the stellar mass is roughly tracing the near infrared
luminosity of a galaxy and the trends are similar to the total mass case.

Thus it appears that the oft-stated claim that galaxy-galaxy lensing probes
the profile of dark matter halos around galaxies is not true in detail.

\section{Strong lensing and external shear} \label{sec:strong}

The angular structure in the gravitational field near a multiple-image
gravitational lens arises from three components: the intrinsic structure of
the primary lens galaxy, local tidal shears generated by other halos and
structures correlated with the primary lens, and the accumulated weak or
large-scale structure (LSS) shear along the ray between the observer and
the source.  
In most circumstances, the LSS contribution can be modeled as an additional
tidal shear near the primary lens
(Kovner~\cite{Kovner87}, Barkana~\cite{Barkana96}).
Simple models for the two sources of tidal perturbations
(Kochanek \& Apostolakis~\cite{Kochanek88};
Keeton, Kochanek \& Seljak~\cite{Keeton97})
suggest that the shear from correlated structures is more important than
the shear from LSS.  Keeton et al.~(\cite{Keeton97}) also found that models
for all four-image lenses, whose geometry makes the models very sensitive to
the angular structure of the gravitational field, show dramatic improvements
when the model has two axes for the angular structure of the gravitational
field.  These two axes presumably arise from the major axis of the primary
lens and the major axis of the combined tidal shear contributions.
Kochanek~(\cite{KocConf}) has shown that one angular component is clearly
aligned with the primary lens galaxy, while the other has the amplitude
expected from tidal perturbations.

While standard weak lensing methods can be used to calculate the statistical
properties of the LSS shear contribution, they are not well suited to
estimating the contribution from structure correlated with the primary lens.
If our simulation is correctly modeling the sites of galaxy formation,
however, it is ideal for exploring the correlations between the shear
generated by the density distribution and estimates of the shear based on
the virialized halos which we can observe as galaxies.
Here we begin this exploration by posing several questions, focusing on the
properties of the correlated shear.
Answers to these questions can be of practical use in understanding
gravitational lenses from observations and models.
First, what is the amplitude and distribution of the shear perturbations
generated by structure correlated with the primary lens?
Second, what is the physical scale on which the shear is typically generated?
Third, how does the total shear correlate in direction and amplitude with the
distribution of galaxies?  

We begin by expanding the potential, projected along a randomly chosen
line-of-sight, in a Fourier series (the projection and Fourier expansion
don't ``commute'').
Of particular interest is the ``shear'' or quadrupole moment of the projected
potential, which we define as the coefficient of the $R^2$ term near the
origin, or
\begin{equation}
  \Phi_2 \equiv
  -G \int d^2R\ {\Sigma(R,\theta)\over R^2}\ e^{2i\theta}
  \qquad .
\label{eqn:gamma2}
\end{equation}
where $R$ is the 2D projected distance.
We obtain the shear by rewriting the potential in terms of the lensing
potential $\psi$, which satisfies $\nabla^2\psi=2\Sigma/\Sigma_{\rm crit}$,
and then taking the absolute value
\begin{equation}
  \gamma \equiv \left| \psi_2\right|
  = \left| \Sigma_{\rm crit}^{-1}{\Phi_2\over 2\pi G} \right|
\end{equation}
where
\begin{equation}
  \Sigma_{\rm crit}\equiv {c^2\over 4\pi G}{D_{LS}\over D_SD_L}
  \qquad .
\label{eqn:sigmacrit}
\end{equation}
For a source at $z=1$ and a lens at $z=0.5$ the {\it comoving\/} critical
density is $2\times 10^{15}\,h^{-1}M_\odot/(h^{-1}{\rm Mpc})^2$.
Throughout we shall quote $\Sigma_{\rm crit}\psi_2$ which can be scaled to
any given lens redshift using Eq.~(\ref{eqn:sigmacrit}).

In Eq.~(\ref{eqn:gamma2}) we need to define the region of integration.  We
are not interested in the contribution to this shear from the galaxy itself,
which is in any case difficult for us to resolve.  Nor are we interested in
the contribution from uncorrelated large-scale structure along the
line-of-sight.  Thus we shall compute the shear within a shell extending from
$r_{\rm min}$ to $r_{\rm max}$, where $r$ denotes a 3D distance.
For $r_{\rm max}$ we choose roughly twice the correlation length of the
galaxy-mass cross correlation function or $2h^{-1}$Mpc.
In fact the shear converges well within this radius and the results for
$r_{\rm max}=1h^{-1}$Mpc are almost identical.
Since the integral converges rapidly the change of the chord length at
large-$r$ doesn't affect our results.
For $r_{\rm min}$ we choose two scales which bracket the reasonable range.
At the low end we choose $r_{\rm min}=50h^{-1}$kpc, just outside the region
of baryon domination.  At the high end we choose $r_{\rm min}=200h^{-1}$kpc,
roughly twice the virial radius.  The amplitude of our results are quite
sensitive to this choice, as the shear is dominated by nearby structures.
Finally we also want to exclude matter (and galaxies) which would, in
projection, lie close to the Einstein radius of the galaxy.  For this reason
we exclude any matter or galaxies with $R<10h^{-1}$kpc.

Our particle-based estimator for ${\mathbf\gamma}$ is thus
\begin{equation}
  \Sigma_{\rm crit}\gamma \simeq
  {1\over 2\pi}\left| \sum_j {m_j\over R_j^2}\ e^{2i\theta_j} \right|
\end{equation}
where $\theta_j$ and $R_j$ are the projected coordinates of particle $j$ and
the sum is over all particles in a shell, centered on the galaxy, with
$r_{\rm min}<r_j<r_{\rm max}$ and $R_j>10h^{-1}$kpc.

\begin{figure}
\begin{center}
\resizebox{3.5in}{!}{\includegraphics{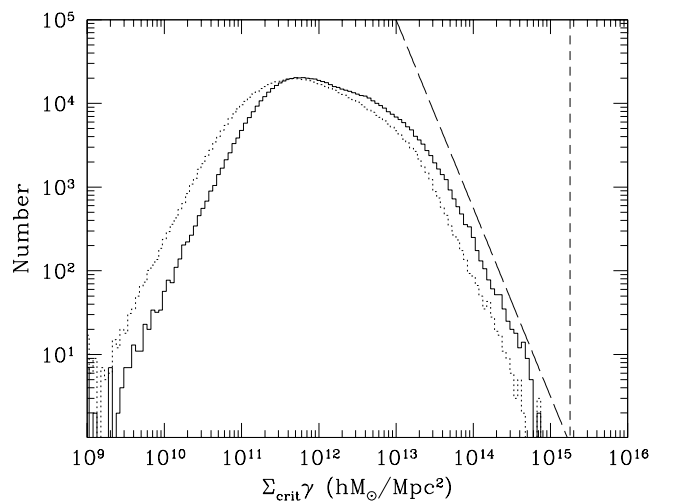}}
\end{center}
\caption{The distribution of $\Sigma_{\rm crit}\gamma$ around galaxies more
massive than $10^{10}h^{-1}M_\odot$.  The solid line marks the shear computed
within $50h^{-1}{\rm kpc}<r_j<2h^{-1}$Mpc and $R_j>10h^{-1}$kpc and the dotted
line $200h^{-1}{\rm kpc}<r_j<2h^{-1}$Mpc.
Varying $r_{\rm max}$ makes almost no difference to the distribution.
The vertical dashed line marks $\Sigma_{\rm crit}$ for a source at $z=1$ and
the long-dashed line has slope $-9/4$ (see text).}
\label{fig:shearg}
\end{figure}

The distribution in amplitude of the shear, $\Sigma_{\rm crit}\gamma$, is
shown in Fig.~\ref{fig:shearg}.  For $\gamma>10^{-2}$ the distribution is
quite well fit by a power-law, with the amplitude somewhat dependent on
the value of $r_{\rm min}$ we choose.  Towards lower values of the shear we
have a roll-off, as required when the Universe becomes optically thick.
Comparing the amplitude of the shear generated with
$r_{\rm min}=50h^{-1}$kpc to that with $r_{\rm min}=200h^{-1}$kpc we see
that much of the shear is generated close to the lens.
The slope of the distribution is close to the prediction of
Keeton et al.~(\cite{Keeton97}), who modelled the ``external'' shear as due
to singular isothermal spheres of fixed radius (for which $\gamma\sim R^{-1}$)
distributed according to a power-law correlation function
$\xi(r)\sim r^{-\chi}$ with $\chi=7/4$.  In this model $P(>\gamma)$ is just
$P(<R)$ so $dP/d\gamma\propto\gamma^{\chi-4}=\gamma^{-9/4}$ (shown as the
dashed line in Fig.~\ref{fig:shearg}).

\begin{figure}
\begin{center}
\resizebox{3.5in}{!}{\includegraphics{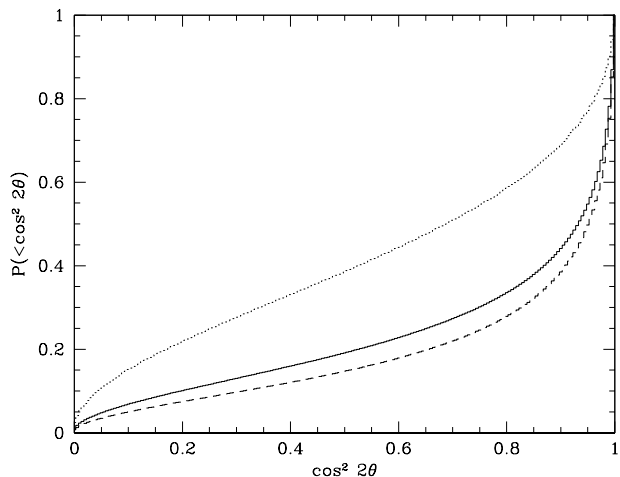}}
\end{center}
\caption{The distribution of $\cos^22\theta$, the angle between the shear
determined using all of the mass and the shear determined using galaxies
(weighted by their mass).  The solid line gives the misalignment using all
of the galaxies in the $r_{\rm min}=50h^{-1}$kpc shell to estimate
$\gamma_{\rm gal}$ and the dashed line shows the effect of increasing
$r_{\rm min}$ to $200h^{-1}$kpc.  In each case the misalignment taking just
the closest massive ($M_{500}>10^{12}h^{-1}M_\odot$) galaxy is almost
indistinguishable.  The dotted line shows the misalignment using just the
closest galaxy above $10^{10}h^{-1}M_\odot$ for the case
$r_{\rm min}=50h^{-1}$kpc.}
\label{fig:shear_dir}
\end{figure}

Another interesting question is how the shear computed using all of the mass
compares to that obtained by using only the galaxies.
We recomputed the shear above using all galaxies in the annulus with
$M_{500}>10^{10}h^{-1}M_\odot$.
We find that the direction is reasonably well reproduced, with the cosine of
twice the misalignment angle
\begin{equation}
  \cos^2 2\theta = { \left| \gamma_{\rm mass}^{*}\gamma_{\rm gal} \right|^2
                     \over
                     \left| \gamma_{\rm mass}\right|^2
                     \left| \gamma_{\rm gal} \right|^2 }
\end{equation}
sharply peaked near 1 (Fig.~\ref{fig:shear_dir}).
It is also of interest to ask how our result is changed if we take only the
nearest galaxy.  Here we find that while the shear is still strongly peaked
near $\cos^22\theta=1$, it is less strongly peaked than if we use all of the
galaxies (Fig.~\ref{fig:shear_dir}).  A much better indicator is using the
nearest {\it massive\/} galaxy (i.e.~galaxy above $10^{12}h^{-1}M_\odot$) in
which case the alignment is almost as good as using all of the galaxies in
the shell.
In comparing to observations of course it is important to include the shear
coming from uncorrelated large-scale structure along the line-of-sight which
may cause a swing in the angle away from the direction of the nearest galaxy.

Finally, scaling our results to the same mean mass density we find that the
amplitude ratio
\begin{equation}
  {\cal R} = { \left| \gamma_{\rm mass} \right| \over
               \left| \gamma_{\rm gal} \right| }
\end{equation}
has a large scatter, with values covering several decades.
This can be understood physically by recalling that the galaxies roughly trace
the position of mass concentrations, but the amount of mass in the galaxies
can be only a small fraction of the total mass in a given halo.  This fraction
typically decreases as the halo mass increases.


\section{Conclusions} \label{sec:conclusions}

We have presented predictions for the distribution of mass around sites of
galaxy formation using a hydrodynamic simulation of structure formation which
includes cooling, star-formation and feedback.
In such simulations galaxies can be easily identified as dense knots of
gas which stand out strongly from the background.  Thus these simulations can
be used to predict, from first principles, the galaxy auto-correlation
function, the galaxy-mass cross correlation function (which is the key
ingredient in galaxy-galaxy lensing studies) and the mass auto-correlation
function.

While the dark matter clustering agrees very well with fitting functions
we find that the evolution of the galaxy `bias' is non-trivial and
non-monotonic and that the bias is increasingly `stochastic' to small
scales.  The galaxy-mass cross-correlation function is approximately the
geometric mean of the galaxy-galaxy and mass-mass auto-correlation functions
on scales above a few hundred kpc but there is more structure below 100kpc.

We find that halos in our simulation have approximately NFW forms, with a
large scatter about the mean profile.
Relating the galaxy-mass cross-correlation to the profile of an ``average''
halo is however fraught with difficulties.  Neither the mean mass, the
profile or the rotation curve of a typical halo is well reproduced by
interpreting $\xi_{\rm gm}(r)$ as a galaxy profile.
This casts some doubt on the ability of galaxy-galaxy lensing to determine
the halo properties or mass-to-light ratios of ``typical'' galaxies.

Finally we have looked at the distribution of shear around galaxies which
could be strong gravitational lenses.  We found that the cosmologically
relevant distribution of shears is well approximated by a power-law.
The slope of this power-law is in good agreement with a model which assumes
all galaxies are singular isothermal spheres with a power-law correlation
function and that the shear is dominated by the nearest neighbour.
The distribution rolls over at $\gamma\sim 10^{-2}$ with a peak roughly an
order of magnitude below this.
The direction of the shear is well reproduced by assuming the galaxies trace
the mass, or that the shear is dominated by the nearest massive galaxy.
The distribution of the magnitude of the shear is however quite broad.

\section*{Acknowledgments}

M.W.~would like to thank Chris Kochanek for numerous helpful conversations
about the material in \S\ref{sec:strong} and a careful reading of the
manuscript.
This work was supported in part by the Alfred P. Sloan Foundation and the
National Science Foundation, through grants PHY-0096151, ACI96-19019 and
AST-9803137.


\begin{thebibliography}{99}
\bibitem[2001]{AMUS}
  Adami C., Mazure A., Ulmer M.P., Savine C., 2001, preprint [astro-ph/0102133]
\bibitem[2000]{BRE}
  Bacon D., Refregier A., Ellis R., 2000, \mnras, 318, 625
\bibitem[1996]{Barkana96}
  Barkana, R., 1996, \apj, 468, 17
\bibitem[1996]{BraBlaSma}
  Brainerd T., Blandford R.D., Smail I., 1996, \apj, 466, 623
\bibitem[2001]{2dF-Kband}
  Cole S., et al., 2001, preprint [astro-ph/0012429]
\bibitem[1999]{DHKW}
  Dav\'e, R., Hernquist, L., Katz, N., Weinberg, D.H., 1999, ApJ, 511, 521
\bibitem[1996]{DelTys}
  Dell'Antonio I.P., Tyson J.A., 1996, \apj, 473, L17
\bibitem[2000]{SDSS}
  Fischer P., et al., 2000, \aj, 120, 1198 [astro-ph/9912119]
\bibitem[1996]{GCIR}
  Griffiths R.E., Casertano S., Im M., Ratnatunga K.U., 1996, \mnras, 282, 1159
\bibitem[2000]{GuzSel}
  Guzik J., Seljak U., 2000, preprint [astro-ph/0007067]
\bibitem[1996]{HaaMad}
  Haardt F., Madau P., 1996, ApJ, 461, 20
\bibitem[1989]{HerKatz}
  Hernquist, L. \& Katz, N., 1989, ApJS, 70, 419
\bibitem[1988]{HocEas}
  Hockney R.W., Eastwood J.W., 1988, Computer Simulation Using Particles,
  Adam Hilger, Bristol
\bibitem[1998]{HGDK}
  Hudson M.J., Gwyn S.D.J., Dahle H., Kaiser N., 1998, \apj, 503, 531
\bibitem[1999]{HulPha}
  Hultman J., Pharasyn A., 1999, A\&A, 347, 769
\bibitem[1995]{JaiMoWhi}
  Jain B., Mo H.J., White S.D.M., 1995, \mnras, 276, L25
\bibitem[1998]{Jen98}
 Jenkins A., Frenk C.S.,  Pearce F.R., Thomas P.A., Colberg J.M., White S.D.M.,
 Couchman H.M.P., Peacock J.A., Efstathiou G., Nelson A.H., 1998, \apj, 499, 20
\bibitem[2000]{JFWCCEY}
 Jenkins A., Frenk C.S., White S.D.M., Colberg J.M., Cole S., Evrard A.E.,
 Yoshida N., 2000, MNRAS, in press [astro-ph/0005260]
\bibitem[1992]{Kai92}
  Kaiser N., 1992, \apj, 388, 272
\bibitem[2000]{KWL}
  Kaiser N., Wilson G., Lupino G., 2000, preprint [astro-ph/0003338]
\bibitem[1992]{KHW92}
  Katz N., Hernquist L., Weinberg D.H., 1992, ApJ, 399, L109
\bibitem[1999]{KHW99}
  Katz N., Hernquist L., Weinberg D.H., 1999, ApJ, 523, 463
\bibitem[1996]{KatWeiHer}
  Katz N., Weinberg D.H., Hernquist L., 1996, ApJS, 105, 19
\bibitem[1997]{Keeton97}
  Keeton, C.R., Kochanek, C.S., \& Seljak, U., 1997, \apj, 482, 604
\bibitem[2001]{KocConf}
  Kochanek C.S., 2001, in ``The shapes of galaxies and their halos'', Yale
  Cosmology Workshop, May 28-30, New Haven CT, ed. Priya Natarajan,
  World Scientific.
\bibitem[1988]{Kochanek88}
  Kochanek, C.S. \& Apostolakis, J., 1988, \mnras, 235, 1073
\bibitem[1987]{Kovner87}
  Kovner, I., 1987, \apj, 316, 52
\bibitem[2001]{Maoli}
  Maoli R., et al., 2001, A\&A, in press [astro-ph/0011251]
\bibitem[1999]{MeiWhi}
  Meiksin A., White M., 1999, \mnras, 308, 1179
\bibitem[2001]{MetWhiLok}
  Metzler C., White M., Loken C., 2001, \apj, 547, 560
\bibitem[1998]{NKSE}
  Natarajan P., Kneib J., Smail I., Ellis R.S., 1998, \apj, 499, 600
\bibitem[1996]{NFW}
  Navarro J., Frenk C.S., White S.D.M., 1996, ApJ, 462, 563
\bibitem[1995]{OstSte}
  Ostriker J., Steinhardt P.J., 1995, Nature, 377, 600
\bibitem[1996]{PD96}
  Peacock J.A., Dodds S.J., 1996, \mnras, 280, L19
\bibitem[1997]{Rauchetal}
  Rauch, M., Miralda-Escude, J., Sargent, W.L.W., Barlow, T.A.,
  Weinberg, D.H., Hernquist, L., Katz, N., Cen, R. \& Ostriker, J.P.,
  1997, ApJ, 489, 7
\bibitem[2001]{RRG}
  Rhodes J. Refregier A., Groth E., 2001, preprint [astro-ph/0101213]
\bibitem[2000]{SSHJ}
  Scoccimarro R., Sheth R., Hui L., Jain B., 2000, preprint
  [astro-ph/0006319]
\bibitem[2000]{Sel}
  Seljak U., 2000, preprint [astro-ph/0001493]
\bibitem[2000]{SWTK}
  Springel V., White S.D.M., Tormen G., Kauffman G., 2000, preprint
  [astro-ph/0012055]
\bibitem[2001]{SprYosWhi}
  Springel V., Yoshida N., White S.D.M., 2001, New Astronomy 6, 79
  [astro-ph/0003162]
\bibitem[1984]{TVJM}
  Tyson J.A., Valdes F., Jarvis J.F., Mills A.P. Jr, 1984, \apj, 281, L59
\bibitem[2000]{Ludo}
  van Waerbeke L.V., et al., 2000, A\&A, 358, 30
\bibitem[2001]{HaloMass}
  White M., 2001, A\&A, 367, 27 [astro-ph/0011495]
\bibitem[2001]{WhiHerSpr}
  White M., Hernquist L., Springel V., 2001, \apj, 550, 129 [astro-ph/0012518]
\bibitem[2000]{WKLC}
  Wilson G., Kaiser N., Lupino G.A., Cowie L.L., 2000, preprint 
  [astro-ph/0008504]
\bibitem[2000]{WTKDB}
  Wittman D.M., et al., 2000, Nature, 405, 143
\bibitem[1997]{YepKatKhoKly}
  Yepes G., Kates R., Khokhlov A., Klypin A., 1997, MNRAS, 284, 235
\end{thebibliography}
\end{document}